\documentclass[a4paper,11pt]{article}
\pdfoutput=1 

\usepackage{jheppub} 

\usepackage{subfigure}
\usepackage[T1]{fontenc} 

\usepackage{mathrsfs}
\usepackage{amsmath}

\usepackage{hyperref}

\usepackage{graphicx}
\usepackage{dcolumn}
\usepackage{bm}
\usepackage[normalem]{ulem}

\def\0{\mbox{\boldmath$\displaystyle\boldsymbol{0}$}}

\newcommand{\dual}[1]{\overset{\:{}^{^{{{\neg}}}}}{\smash[t]{#1}}} 

\newcommand{\dualn}[1]{\overset{\:{}^{^{^{{{\neg}}}}}}{\smash[t]{#1}}} 

\title{\boldmath Elko as an inflaton candidate}

\author[a]{Xinglong Chen,}
\affiliation[a]{Center for theoretical physics, College of Physics, Sichuan University,\\Chengdu, 610064, China} 
\emailAdd{chenxinglong@stu.scu.edu.cn}  
\author[a]{Cheng-Yang Lee,}
\emailAdd{cylee@scu.edu.cn}
\author[b]{Yanjiao Ma,}
\affiliation[b]{The Hong Kong University of Science and Technology,\\Clear Water Bay, Kowloon, Hong Kong, P.R.China} 
\emailAdd{ymaby@connect.ust.hk}
\author[c]{Haomin Rao,}
\affiliation[c]{School of Intelligent Engineering, Shaoguan University, Shaoguan 512005, China} 
\emailAdd{rhm137@mail.ustc.edu.cn}
\author[b]{Wenqi Yu}
\emailAdd{wyuaz@connect.ust.hk}
\author[d]{and Siyi Zhou}
\emailAdd{siyi@cqu.edu.cn} 
\affiliation[d]{Department of Physics and Chongqing Key Laboratory for Strongly Coupled Physics,\\
Chongqing University, Chongqing 401331, China}

\abstract{
Elko is a spin-half fermion with a two-fold Wigner degeneracy and Klein-Gordon dynamics. In this paper, we show that in a spatially flat FLRW space-time, slow-roll inflation can be initiated by the homogeneous Elko fields. The inflaton is a composite scalar field obtained by contracting the spinor field with its dual. This is possible because the background evolution as described by the Friedmann equation is completely determined by the scalar field. This approach has the advantage that we do not need to specify the initial conditions for every component of the spinor fields. We derive the equation of motion for the inflaton and also show that this solution is an attractor. 
Finally, we examine the slow-roll parameters and the power-spectrum, showing that obtaining a behavior in agreement with observational requirements is hard to be obtained, unless one uses more complicated potentials, which may act  a limitation of Elko inflation.
}

\begin{document} 
\maketitle
\flushbottom

\section{Introduction}

Inflation is a theory of rapid spatial expansion of the early universe. It is the leading paradigm describing the early universe cosmology because it resolves the flatness, horizon, and magnetic monopole problems~\cite{Guth:1980zm, Barrow:1981pa, Linde:1983gd, Linde:1981mu}. The slow-roll inflation driven by a real scalar field is one of the simplest models. It predicts an almost scale-invariant power spectrum which is in agreement with observations on the cosmological background radiation~\cite{Mukhanov:1990me}. Heavy fields may be present during inflation. Their presence leads to interesting oscillating features in the squeezed limit of non-Gaussianities as is the case in the cosmological collider physics or quasi-single field inflation~\cite{Chen:2010xka, Chen:2009we, Chen:2009zp, Chen:2012ge, Noumi:2012vr, Arkani-Hamed:2015bza, Meerburg:2016zdz, Chen:2016uwp, Tong:2017iat, Tong:2018tqf, Alexander:2019vtb, Lu:2019tjj, Hook:2019zxa, Wang:2019gbi, Wang:2020uic, Wang:2020ioa, Liu:2019fag, Lu:2021wxu, Tong:2022cdz, Tong:2021wai}.  
Nowadays, there are many inflationary models. Apart slow-roll scalar inflation~\cite{Martin:2013tda},
one may also think about inflation driven by spinor~\cite{Kinney:1995xv, Iso:2014gka, Pereira:2017efk, Kumar:2018rrl, Benisty:2019jqz, Shokri:2021aum, Gredat:2008qf, Shankaranarayanan:2009sz} or vector fields~\cite{Golovnev:2008cf, Koivisto:2008xf, Chiba:2008eh, Koh:2009ne, Setare:2013kja, Darabi:2014aaa}. 


In this work, we study the theory of Elko inflation. Elko is a massive spin-half fermionic field in the $\left(\frac{1}{2},0\right)\oplus\left(0,\frac{1}{2}\right)$ representation with double Wigner degeneracy~\cite{Ahluwalia:2004ab, Ahluwalia:2004sz, Ahluwalia:2008xi, Ahluwalia:2009rh, Ahluwalia:2010zn, Lee:2012td, Lee:2014opa, Lee:2015sqj, Ahluwalia:2016rwl, Lee:2019fni, Ahluwalia:2019etz, Ahluwalia:2022ttu, Ahluwalia:2022yvk, Ahluwalia:2023slc}. Double Wigner degeneracy, refers to the fact that the fermionic states belong to the irreducible unitary representation of the extended Poincar\'{e} group. These fermionic states have four degrees of freedom, two coming from spin projections and two coming from Wigner degeneracy $n=1,-1$. The latter is defined by the discrete symmetry transformations where parity and time-reversal map a single state to a superposition of two states labeled by $n=1$ and $n=-1$.\footnote{For the Standard Model fermions, parity and time-reversal transformations map a single state to itself with momentum and spin-projections altered.}

 One is justified to ask the question \textit{why use a spinor field to drive inflation when a scalar field suffices?} It is true that scalar inflationary models would be simpler than spinor ones. However, our main purpose here is not to propose another inflationary model. Instead, this work is part of an ongoing program to systematically study Elko in cosmology. So in our opinion, investigating whether Elko can drive inflation is a topic worthy of investigation as a part of the ongoing program. Therefore, a fair attitude towards this work is that we have demonstrated that Elko can in principle, be used to drive inflation. Moreover, choosing Elko to drive inflation has many advantages, for example, its status as a candidate for dark matter means it could not only drive inflation but may also explain dark matter after inflation ends. 


The idea of Elko inflation is not new~\cite{Boehmer:2006qq,Boehmer:2007dh,Boehmer:2008rz,Boehmer:2008ah,Boehmer:2009aw,Boehmer:2010ma,Basak:2012sn,Sadjadi:2012xyd,HoffdaSilva:2014tth,Pereira:2014wta,S:2014woy,Pereira:2014pqa,Chang:2015ufa,Pereira:2016emd,Pereira:2016eez,Pereira:2017efk, Basak_2015}. But prior to~\cite{Ahluwalia:2022yvk, Ahluwalia:2023slc}, the correct degrees of freedom were not known. As a result, these models are non-local and not Lorentz covariant in Minkowski space-time. The realization that Elko are fermionic fields with double Wigner degeneracy resolves the problems of non-locality and Lorentz violation in Minkowski space-time. In de Sitter space-time, it is shown that Elko can be consistently quantized where the quantum fields satisfy the canonical anti-commutations relations~\cite{Lee:2024sbg}. 

Here we show that Elko can act as an inflaton using the methods in~\cite{Benisty:2019jqz}. Firstly, to the zeroth order in perturbation, we take the Elko field $\lambda$ and its dual $\dual{\lambda}$ to be homogeneous. Their perturbations will be studied in future works. Next, we show that in the spatially flat FLRW space-time, all components of the Elko energy-momentum tensor can be expressed in terms of $\phi$. Therefore, by solving the equation of motion for $\phi$ (which can be derived from the equations of motion for $\lambda$ and $\dual{\lambda}$) and the Friedmann equations, we can determine the evolution of the Hubble parameter and the scale factor. 

By choosing the appropriate parameters for the Elko potential and numerically solving for $\phi$, we can obtain slow-roll inflation. Compared to the existing literature, the advantage of our approach is that the equation of motion for $\phi$ is a third-order differential equation so we only need to specify three initial conditions for $\phi$, $\dot{\phi}$ and $\ddot{\phi}$. 
If we directly solve the equations of motion for the spinor fields which are second-order differential equations, we would have to specify a total of sixteen initial conditions for $\lambda_{\ell},\dual{\lambda}_{\ell},\dot{\lambda}_{\ell},\dot{\dual{\lambda}_{\ell}}$ where $\ell=1,\cdots,4$.

The paper is organized as follows. In sec.~\ref{elko}, we derive the Elko energy-momentum tensor in curved space-time. In sec.~\ref{elkoFLRW}, we study in the spatially flat FLRW space-time. We derive the equations of motion of $\phi$ and express the Elko energy-momentum tensor in terms of $\phi$. In sec.~\ref{attractor}, we find that with the appropriate choice of parameters, Elko can act as an inflaton. Moreover, the inflation solution is also an attractor. We give a brief conclusion and outlook in sec.~\ref{conc}. In app.~\ref{Example}, we explain why the third-order differential equation for $\phi$ does not give rise to Ostrogradsky ghosts.



\section{Set up}\label{elko}


In Minkowski space-time, the equation of motion for Elko is the spinorial Klein-Gordon
equation. In curved space-time, the equation of motion for Elko is
\begin{align}\label{Elkoeom0}
\gamma^{\mu}\nabla_{\mu}(\gamma^{\nu}\nabla_{\nu}\lambda)-\frac{\partial V}{\partial\dualn{\lambda}}=0\,,
\end{align}
where $\lambda$, $\dual{\lambda}$ are the Elko field and its dual field and $V=V(\dual{\lambda}\lambda)$ is the interacting potential. In this paper, we work within the framework of general relativity. Therefore, as explained in~\cite{Lee:2024sbg}, the complete Einstein-Elko action is
\begin{align}
    S&=S_{\text{GR}}+S_{\text{Elko}}\nonumber\\
    &=\int d^{4}x\sqrt{-g}\left[\frac{1}{2}M^{2}_{p}R
    -g^{\mu \nu}(\nabla_\mu \dual{\lambda})(\nabla_{\nu}\lambda)-\frac{1}{4}R\dual{\lambda}\lambda-V(\dual{\lambda}\lambda)\right]\,,\label{eq:e_e_action}
\end{align}
where $M_{p}$ is the reduced Planck mass, $g^{\mu\nu}$ is the metric and $R$ is the Ricci scalar.
The quadratic term $\frac{1}{4}R\dual{\lambda}\lambda$ is necessary to ensure that Elko satisfies the spinorial Klein-Gordon equation (\ref{Elkoeom0}). Since our action~(\ref{eq:e_e_action}) is differs from the previous works, it is necessary to revisit the analysis that leads to Elko inflation.

In the analysis of spinor fields in curved spacetime, it is convenient to use the tetrad formalism.
The tetrads and the metric satisfy the following relationship
\begin{align}
\begin{aligned} g_{\mu\nu}&=e^{a}_{~\mu} e^{b}_{~\nu} \eta_{ab}\,, \\ \eta_{ab} & =e_{a}^{~\mu} e_{b}^{~\nu} g_{\mu\nu}\,,\end{aligned}
\end{align}
where the local and global coordinates are labeled by the Latin ($a,b,\cdots$) and Greek ($\mu,\nu,\cdots$) alphabets respectively. 
Both the metric and the tetrad are compatible with the connection,
\begin{equation}
    \nabla_{\sigma}g_{\mu\nu}=0\,,\quad\nabla_{\sigma}{e^{a}}_{\nu}=0\,.\label{eq:dg}
\end{equation}
The Dirac matrices satisfy the anti-commutation reation $\{\gamma^{\mu},\gamma^{\nu}\}=2g^{\mu\nu}\mathbb{I}_{4}$ and their indices can be raised and lowered by the tetrad
\begin{align}
\gamma^\mu =e_a^{~\mu} \gamma^a\,,\quad
\gamma_\mu =e^b_{~\mu} \gamma_b\,.
\end{align}
The result of the covariant derivative acting on Elko and its adjoint are
\begin{align}
\nabla_\mu\dual{\lambda}&=\partial_{\mu}\dual{\lambda}-\dual{\lambda}\Gamma_{\mu}\,, \label{eq:cov1}\\    
\nabla_\mu \lambda&=\partial_\mu \lambda+\Gamma_\mu \lambda\,,\label{eq:cov2}
\end{align}
where 
\begin{equation}
\Gamma_\mu =\frac{i}{2}{\omega_{\mu}}^{ab}\Sigma_{ab}\,,
\end{equation}
and
\begin{equation}
  \Sigma^{ab}=-\frac{i}{4}[\gamma^{a},\gamma^{b}]  
\end{equation}
is the Lorentz generator of the spinor field, and ${\omega_{\mu}}^{ab}$ is the spin connection which has the following relationship with the Christoffel symbol $\Gamma^{\nu}_{~\mu\sigma}$ 
\begin{equation}\label{relation1}
{\omega_{\mu}}^{ab}={e^{a}}_{\nu}\left(\partial_{\mu}e^{b\nu}+e^{b\sigma}\Gamma^{\nu}_{~\mu\sigma}\right)\,.
\end{equation}

We now derive the Elko energy-momentum tensor using $S_{\text{Elko}}$. Firstly, we vary the tetrad $\delta e^{a}{}_{\mu}$ and the spin connections $\delta\omega_{\mu ab}$ independently
\begin{align}\label{variation1}
 \delta S_{\text{Elko}}=\int d^{4}x\sqrt{-g}\, \left( \sigma^{\mu\nu}e_{a\mu}\delta e^{a}_{~\nu}+\tau^{ab\mu}\delta\omega_{\mu ab}\right)\,,
\end{align}
where 
\begin{align}
\sigma^{\mu\nu}=&2\nabla^{(\mu}\dual{\lambda}\nabla^{\nu)}\lambda-g^{\mu\nu}\left[(\nabla^\gamma \dual{\lambda})(\nabla_{\gamma}\lambda)+\frac{1}{4}R\dual{\lambda}\lambda+V(\dual{\lambda}\lambda)\right]\nonumber\\
&+\frac{1}{2}\left[R^{\mu\nu}\dual{\lambda}\lambda+(g^{\mu\nu}\nabla^{\sigma}\nabla_{\sigma}-\nabla^{\mu}\nabla^{\nu})\dual{\lambda}\lambda\right]\,,\label{eq:sigma}
\end{align}
and
\begin{align}
\tau^{ab\mu}&=\frac{i}{2}\left(\dual{\lambda}\Sigma^{ab}\nabla^{\mu}\lambda-\nabla^{\mu}\dual{\lambda}\Sigma^{ab}\lambda\right)\,.
\end{align}
The parenthesis on the indices is defined as
\begin{equation}
    T_{(\mu_1 \mu_2\cdot\cdot\cdot\mu_n)\rho}^{\sigma} = \frac{1}{n!}\left[T_{\mu_1\mu_2\cdot\cdot\cdot\mu_n \rho}^{\sigma} +\left(\text{permutations of }\mu_1 \mu_2\cdot\cdot\cdot\mu_n\right)\right]\,.
\end{equation}
In the above derivation for~(\ref{eq:sigma}), we have used
\begin{align}
\delta R&=-R^{\mu\nu}\delta g_{\mu\nu}+(\nabla^{\mu}\nabla^{\nu}-g^{\mu\nu}\nabla^{\sigma}\nabla_{\sigma})\delta g_{\mu\nu}\nonumber\\
&=-2R^{\mu\nu}e_{a(\mu}\delta e^{a}{}_{\nu)}+
2\left(\nabla^{\mu}\nabla^{\nu}-g^{\mu\nu}\nabla^{\sigma}\nabla_{\sigma}\right)\left[e_{a(\mu}\delta e^{a}{}_{\nu)}\right],
\end{align}
and
\begin{equation}
    \delta g_{\mu\nu}=2e_{a(\mu}\delta e^{a}{}_{\nu)}\,. \label{eq:dg}
\end{equation}
Since the tetrad and spin connection are not independent but satisfy~(\ref{relation1}), the variation of the spin connection can be expressed as the variation of the tetrad
\begin{equation}\label{relation2}
   \tau^{ab\mu}\delta\omega_{\mu ab}=-(\tau^{\mu\nu\rho}+2\tau^{\rho(\mu\nu)})\nabla_{\rho}(e_{a\mu}\delta e^{a}_{~\nu})\,.
\end{equation}
Substituting~(\ref{relation2}) into~(\ref{variation1}), we obtain
\begin{equation}\label{variation2}
    \delta S_{\text{Elko}}=\int d^{4}x\sqrt{-g}\left[\sigma^{\mu\nu}+\nabla_{\rho}(\tau^{\mu\nu\rho}+2\tau^{\rho(\mu\nu)})\right]e_{\alpha\mu}\delta e^{\alpha}_{~\nu}\,.
\end{equation}
Finally using~(\ref{eq:dg}), the Elko energy-momentum tensor is given by
\begin{align}\label{EMTofElko}
    T^{\mu\nu}
    &=\frac{1}{\sqrt{-g}}\frac{\delta S_{\text{Elko}}}{e_{a(\mu}\delta e^{a}{}_{\nu)}}\nonumber\\
    &=\sigma^{(\mu\nu)}+\nabla_{\rho}\left[\tau^{(\mu\nu)\rho}+2\tau^{\rho(\mu\nu)}\right]
    \nonumber\\ & =2\nabla^{(\mu}\dual{\lambda}\nabla^{\nu)}\lambda-g^{\mu\nu}\left[(\nabla^\gamma \dual{\lambda})(\nabla_{\gamma}\lambda)+\frac{1}{4}R\dual{\lambda}\lambda+V(\dual{\lambda}\lambda)\right]+i\nabla_{\sigma}\left[\dual{\lambda}\Sigma^{\sigma(\mu}\nabla^{\nu)}\lambda+\nabla^{(\mu}\dual{\lambda}\Sigma^{\nu)\sigma}\lambda\right]
\nonumber\\ & ~~~+\frac{1}{2}\left[R^{\mu\nu}\dual{\lambda}\lambda+(g^{\mu\nu}\nabla^{\sigma}\nabla_{\sigma}-\nabla^{\mu}\nabla^{\nu})\dual{\lambda}\lambda\right]\,.
\end{align}

\section{Elko in the spatially flat FLRW space-time}\label{elkoFLRW}

We now study Elko in the spatially flat FLRW space-time with the metric
\begin{equation}
    ds^{2}=-dt^{2}+a^{2}(t)\delta_{ij}dx^{i}dx^{j}\,,
\end{equation}
where $a$ is the scale factor. Here, we only consider the zeroth order in perturbation for the Elko field and its dual so they are homogeneous $\lambda=\lambda(t)$, $\dual{\lambda}=\dual{\lambda}(t)$. By solving the equations of motion for Elko and the Friedmann equations, we can determine the background evolution. They are given by
\begin{align}\label{Beom1}
\ddot{\dualn{\lambda}}+3H\dot{\dualn{\lambda}}+\left(\frac{3}{2}\dot{H}+\frac{9}{4}H^{2}\right)\dual{\lambda}+\frac{dV}{d\lambda}&=0\,,\\
\ddot{\lambda}+3H\dot{\lambda}+\left(\frac{3}{2}\dot{H}+\frac{9}{4}H^{2}\right)\lambda+\frac{dV}{d\dualn{\lambda}}&=0\,,\label{Beom2}
\end{align}
and
\begin{align}
   &3H^{2}=M^{-2}_{p}\rho\,, \label{eq:f1}
   \\ & 2\dot{H}+3H^{2}=-M^{-2}_{p}p\,, \label{eq:f2}
\end{align}
where $\dot{f}\equiv df/dt$ and $H=\dot{a}/a$ is the Hubble parameter. In the Friedmann equations, $\rho$ and $p$ are the energy density and pressure of Elko which can be obtained from~(\ref{EMTofElko})
 \begin{equation} 
 T_{00}=\rho\,,~~ T_{0i}=0\,,~~ T_{ij}=a^{2}p\delta_{ij}\,, 
 \end{equation}
where
\begin{align}
   &\rho=\dot{\dualn{\lambda}}\dot{\lambda}+\frac{3}{2}H(\dot{\dualn{\lambda}}\lambda+\dual{\lambda}\dot{\lambda})+\frac{9}{4}H^{2}\dual{\lambda}\lambda+V\,,
   \\ & p=-V+\frac{1}{2}\left(\frac{dV}{d\lambda}\lambda+\dual{\lambda}\frac{dV}{d\dualn{\lambda}}\right)\,. \label{eq:p_E}
\end{align}
They satisfy the continuity equation 
\begin{equation}
\dot{\rho}+3H(\rho+p)=0\,.
\end{equation}

These equations are difficult to solve directly for the following reasons. Both $\lambda$ and $\dual{\lambda}$ are four-component spinor fields so~(\ref{Beom1})-(\ref{Beom2}) constitute eight second-order differential equations. To solve them, we need to specify sixteen initial conditions for $\lambda_{\ell},\dual{\lambda}_{\ell},\dot{\lambda}_{\ell},\dot{\dualn{\lambda}}_{\ell}$ where $\ell=1,\cdots,4$. It is tempting to assume that all components of $\lambda_{\ell}$, $\dual{\lambda}_{\ell}$ have the same initial conditions and the same for their derivatives. In this case, the number of initial conditions would be reduced to two. However, as far as we can see, there does not seem to be a good reason a priori for this assumption.

The main result of this paper is that we can study the background evolution without having to solve for $\lambda$ and $\dual{\lambda}$. Instead, it suffices for us to consider the dynamics of the composite scalar field
\begin{equation}
\phi\equiv\dual{\lambda}\lambda\,.
\end{equation}
The equation of motion for $\phi$ is derived by combining~(\ref{Beom1})-(\ref{Beom2}) to yield
\begin{equation}\label{Beom30}  \dddot{\phi} + 9H \ddot{\phi} + (9\dot{H}+27H^{2}+4V_{\phi}+2V_{\phi\phi}\phi) \dot{\phi} + (3\ddot{H} + 27H \dot{H}+27H^{3} + 12H V_{\phi}) \phi =0\,,
\end{equation}
where
\begin{equation}\nonumber
 V_{\phi}\equiv\frac{dV}{d\phi},\quad
 V_{\phi\phi}\equiv \frac{d^{2}V}{d\phi^{2}}.
\end{equation}
Using the equations of motion for $\lambda$ and $\dual{\lambda}$, $\rho$ and $p$ can be expressed in terms of $\phi$ 
\begin{align}
   \rho&=\frac{1}{2}\ddot{\phi}+3H\dot{\phi}+\left(\frac{3}{2}\dot{H}+\frac{9}{2}H^{2}+V_{\phi}\right)\phi+V\,,\label{eq:rhoe}
   \\ 
   p&=-V+V_{\phi}\phi\,.\label{eq:pe}
\end{align}
Therefore, the background evolution is completely determined by the scalar field $\phi$ that satisfies~(\ref{Beom30}).
It should be noted that using a composite scalar field to replace Elko is only an effective description on the cosmological background. This description no longer holds after considering perturbations. We are not really dealing with a scalar field with higher-order derivatives. It can be seen from the original action (\ref{eq:e_e_action}) that the model avoids higher-order derivatives and therefore does not lead to the Ostrogradsky instability.
We explain this more intuitively with a simple example in Appendix \ref{Example}


\section{Elko inflation is an attractor solution}\label{attractor}

We now explore the possibility of inflation driven by the Elko field. Different from chaotic inflation, it can be seen from (\ref{eq:rhoe}) and (\ref{eq:pe}) that $\dot{\phi}=\ddot{\phi}=0$ does not necessarily mean $p=-\rho$, which is the approximate condition required for inflation. Therefore, we should choose the initial conditions and potential $V$ more carefully. 

First, let us discuss what kind of potential and initial conditions can achieve an approximate de Sitter solution.
In the de Sitter background, that is, $\dot{H}=\ddot{H}=\dot{\phi}=\ddot{\phi}=\dddot{\phi}=0$, 
the equations of motion can be reduced to
\begin{align}
   &3H^{2}=\rho=-p\label{dseq1}
   \\ 
   &\rho=V+(V_{\phi}+\frac{9}{2}H^{2})\phi\label{dseq2}
   \\
   &p=-V+V_{\phi}\phi\,\label{dseq3} ,
   \\
   &(9H^{2}+4V_{\phi})\phi=0\,.\label{dseq4}
\end{align}
There are two types of de Sitter solutions to the above equation: One type of de Sitter solution is $\phi=0$, which automatically satisfies $p=-\rho$ and $H^{2}=V(0)/3$ is a constant. 
Therefore, any potential $V(\phi)$ that satisfies $V(0)>0$ can achieve an approximate de Sitter solution near $\phi\approx 0$.

Another type of de Sitter solution requires that the equation $3V+(4-3\phi)V_{\phi}=0$ has solution $\phi=\phi_0$ ($\phi_0$ may not be unique).  A de Sitter solution can then be achieved at $\phi=\phi_0$, and in this case $H^{2}=-4V_{\phi}(\phi_{0})/9$. Therefore, as long as $3V+(4-3\phi)V_{\phi}=0$ has a solution and the solution $\phi=\phi_0$ satisfies $V_{\phi}(\phi_{0})<0$, an approximate de Sitter solution can be achieved near $\phi\approx \phi_0$.

\begin{figure}[h]
\centering
\includegraphics[width=0.4\textwidth]{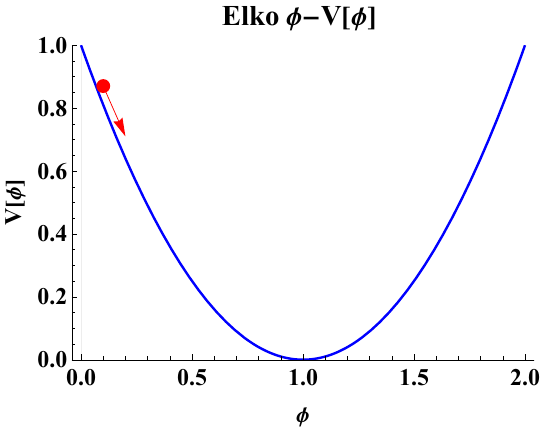}
\caption{The potential energy $V$ with $v_0=M_{p}^{4}, v_1=-2 M_{p}^2$, and $v_2 = 1$, where $V|_{\phi=M_{p}^2} = 0$. The $\phi$ axis is in units of $M^{2}_{p}$.}
\label{Vphi}
\end{figure}

Next, we will use a specific example to illustrate that inflation can be driven by the Elko field.
In the following, we take the
potential $V$ to be a quadratic polynomial of $\phi$ so that it is renormalizable. Therefore,
\begin{equation}
    V=v_0+v_1\phi+v_2\phi^{2}\,. \label{eq:V}
\end{equation}
The coefficient $v_{1}$ can be a constant or it can also be proportional to the Ricci curvature $R$, corresponding to non-minimal coupling. As far as we are concerned, it is acceptable to take $v_{1}$ to be constant because during slow-roll inflation, $R$ is approximately constant. The last term of~(\ref{eq:V}) is a quartic self-interaction of $\lambda$ so $v_2$ is dimensionless. Substituting~(\ref{eq:V}) into~(\ref{eq:p_E}), we obtain
\begin{equation}
    p= -v_0 + v_2 \phi^{2}\,.
\end{equation}
It is worthwhile to note that if $V=v_1 \phi$, then $p=0$ so Elko behaves like dust. 
To make the analysis more specific, we take parameters
$
 v_0 =M_{p}^4,\quad v_1 = -2 M_{p}^2,\quad v_2 = 1,  
$
and the potential is shown in fig.~\ref {Vphi}.

Substituting~(\ref{eq:rhoe})-(\ref{eq:pe}) into (\ref{eq:f1})-(\ref{eq:f2}), we obtain two solutions for $H$
\begin{equation}
H_{\pm}=\frac{ - 6\dot{\phi}\pm A}{9\phi-12 M_{p}^2}\,,
\label{Hphi}
\end{equation}
where 
\begin{equation}
    A = \sqrt{36 \dot{\phi}^2-(9\phi-12 M_{p}^2) \left[\frac{3(V \phi-V_{\phi} \phi^2)}{M_{p}^2} + 4(V_{\phi}\phi + V) +2 \ddot{\phi}\right]}\,,
\end{equation}
and two solutions for $\dot{H}$
\begin{equation}
    \dot{H}_{\pm} = \frac{4V +\ddot{\phi}}{(4 M_{p}^2-3\phi)} - \frac{A^2 \mp 6\dot{\phi} A}{3(4 M_{p}^2-3\phi)^2}\,.
   \label{H1phi}
\end{equation}
To ensure that $H$ is positive and $\dot{H}$ is negative, we need $0 < \phi < \frac{4}{3} M_{p}^2$ (the simplest solution is $V = v_0$), so the solution $H_{+}$ can be ignored. From now onwards $H_{-}\equiv H$.
In this case, the accelerated expansion occurs when the potential energy of the Elko field, $V$ is the dominant contribution to $\rho$.

\begin{figure}[h]
\centering
\subfigure[]{
\label{phiN}
\includegraphics[width=0.4\textwidth]{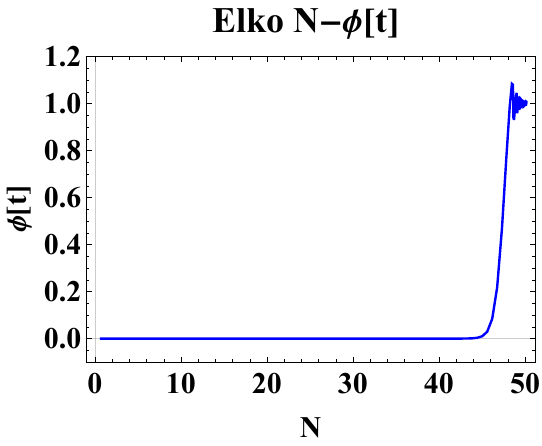}}
\subfigure[]{
\label{HN}
\includegraphics[width=0.4\textwidth]{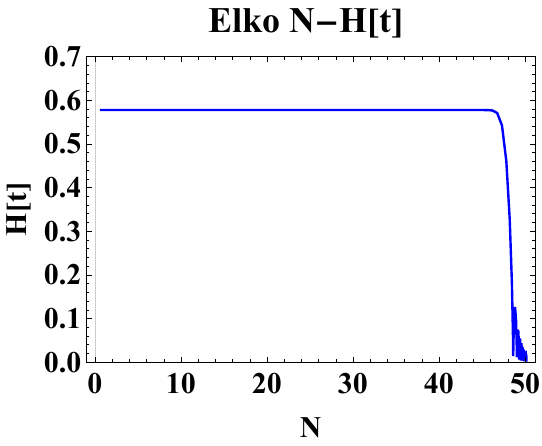}}
\caption{Figure~\ref{phiN} is the evolution of $\phi$ concerning the e-folding number $N$. Figure~\ref{HN} is the evolution of the Hubble parameter $H$ for the $N$, where we can see that the total number of inflationary e-folds exceeds $50$. }
    \label{third solution}
\end{figure}

 Using numerical methods, substituting~(\ref{Hphi}) and~(\ref{H1phi}) into~(\ref{Beom30}), with the initial conditions
\begin{equation}
\phi(0) \sim 0^{+}\,,\quad \dot{\phi}(0) \sim 0^{+}\,,\quad \ddot{\phi}(0) = 0\,,
\end{equation}
we are able to numerically solve the differential equation~(\ref{Beom30}) to obtain the solution as shown in fig.~\ref{third solution}, which results in slow-roll inflation. The number of e-folds depends on the initial condition of $\phi$. Specifically, the closer $\phi(0)$ is to $0$, the larger the number of e-folds.

\begin{figure}[h]
\centering
\includegraphics[width=0.45\textwidth]{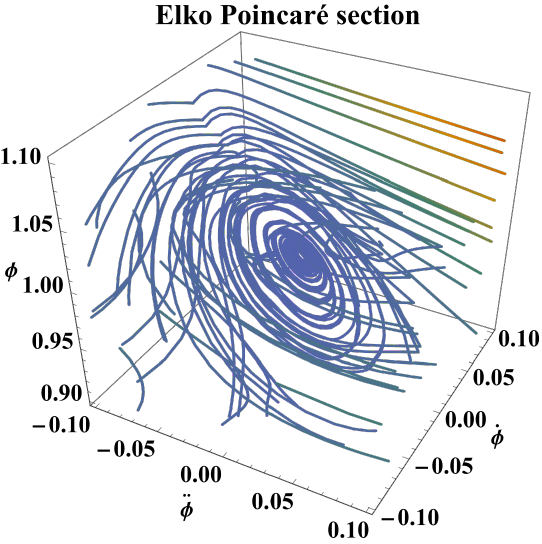}
\caption{The Poincar\'{e} section of the Elko as inflation candidate, where the location of the attractor is $\phi = 1$, $\dot{\phi} = 0$, and $\ddot{\phi} = 0$.}
\label{Poincare section}
\end{figure}

Furthermore, we have solved the equation of motion for $\phi$ and confirmed that the solution is an attractor as shown in fig.~\ref{Poincare section}. When the initial conditions for $\phi(0)$, $\dot{\phi}(0)$, $\ddot{\phi}(0)$ are within a certain range, the final solution is an attractor.

Next, to analyze the power spectrum, we consider the two Hubble slow-roll parameters:
\begin{equation}
    \epsilon=-\frac{\dot{H}}{H^{2}}\,,~\eta=\frac{d \ln \epsilon}{d N}\,,
    \label{espiloneta}
\end{equation}
where $N$ is the number of e-folds. When $\epsilon < 1$, the accelerated expansion starts and it will continue if the change of $\epsilon$ per e-fold is small, that is, $\eta < 1$. 

To match the experimental data, the slow-roll parameters should be on the order of 0.01.
However, the slow-roll parameters given by the above potential are much smaller than 0.01. If we change the initial conditions or the coefficients of the potential to make the slow-roll parameters approximately 0.01, the number of e-folding will be far less than 50. 
To address this issue, we also tested several potentials with different initial conditions. However, we did not find the potential that satisfies the conditions that the slow-roll parameters are of the order 0.01 and, at the same time, generate the appropriate number of e-folding number. 
It seems difficult to realize inflation that is consistent with observations in this model. Perhaps we have not found the appropriate potential and initial conditions.

\section{Conclusion and outlook}\label{conc}

In this paper, we have used the correct Elko action in curved spacetime, and proved that in an Elko-dominated universe, the cosmological background can always be equivalent to being dominated by the composite scalar field $\phi\equiv\dual{\lambda}\lambda$, where $\lambda$ is the Elko field and $\dual{\lambda}$ its dual. Although the equation of motion of the scalar field is a third-order differential equation, it does not lead to ghost instability because the basic field Elko is ghost-free. 

Subsequently, we explored the possibility of inflation driven by the Elko field and have found the general conditions for the de Sitter solution.
Taking a specific potential as an example, 
we solved the third-order differential equation numerically and found that an inflation lasting about 50 e-folding numbers is achievable. Furthermore, from the Poincar\'{e} section, we found that the inflation solution is an attractor.
While the inflationary solution appears promising, a closer look at the power spectrum suggests that the predicted slow-roll parameters may not yet reach the desired magnitude.
We also tried several potentials with different initial conditions, but none of them yielded an inflationary solution that is fully consistent with observational constraints. This may indicate a limitation of the current model or simply reflect that a more suitable potential has yet to be identified.

Although our current analysis has not yielded a fully satisfactory inflationary solution, the theoretical framework offers rich structure and motivates several avenues for future investigation. A natural next step is to examine whether the perturbation of $\phi$ can act as a curvaton field~\cite{Lyth:2001nq, Lyth:2002my, Lyth_2003, Bartolo:2002vf, Sasaki:2006kq}, which is crucial for connecting the model with cosmological observables. As suggested by many curvaton scenarios, such a mechanism could help resolve the preheating problem ~\cite{Bastero-Gil:2003usx, Sainio:2012rp}. With the novel equation of motion found in this work, our model of Elko inflation may predict unique features that can be tested in future experiments. However, a key challenge remains in determining whether the equations of motion for $\lambda$ and $\dual{\lambda}$ can be combined into a single equation governing the perturbation of $\phi$. This is an important question for further study.

In addition to the curvaton possibility discussed above, it is also important to consider the background dynamics that govern the early universe evolution. The current model is based on slow-roll inflation in a quasi–de Sitter (dS) background. However, inflation need not be limited to a quasi-dS background. There are also scenarios such as a power-law inflation~\cite{Galtsov:1997ub, Tsujikawa:2000wc, Feinstein:2002aj, Glavan:2020zne}, where the scale factor is a power-law function of the physical time. Beyond inflation, one may also consider non-inflationary alternatives such as matter bounce ~\cite{Brandenberger:2016vhg, Agrawal:2021rur, Agrawal:2022vdg, Lohakare:2022umj, Agrawal:2022ppe}, ekpyrotic~\cite{Nojiri:2022xdo, Paul:2022mup}, and others~\cite{Gu:2021qwy, Zhu:2015xsa, Zhu:2015owa, Basak:2014qea}. It remains an intriguing question whether Elko can be consistently embedded into these alternative cosmological scenarios.

Finally, as suggested in previous studies~\cite{Pereira:2017efk, Pereira:2017bvq, Pereira:2018hir}, Elko may describe the whole cosmic history. Given the problem of Hubble tension~\cite{DiValentino:2021izs, Perivolaropoulos:2021jda, Botke2023, Sakstein:2019fmf, Niedermann:2020dwg, Dainotti:2023yrk}, there are strong motivations to go beyond the $\Lambda$CDM model. These considerations motivate the exploration of Elko as an alternative to dark energy~\cite{Arun:2017uaw, Meng:2012zza, Wang:2024rus}, with the potential to alleviate the Hubble tension problem.

\section{Acknowledgement}
This work is supported in part by the Natural Science Foundation of China under Grant No.12347101. 
Haomin Rao is supported by the Scientific Research Startup
Foundation of Shaoguan University under Grant No.9900064901.
Yanjiao Ma is supported by the Hong Kong PhD Fellowship Scheme (HKPFS) issued by the Research Grants Council (RGC) of Hong Kong. 

\appendix
\section{Composite scalar field with third-order equation does not imply ghost}~\label{Example}

In sec.~\ref{elkoFLRW}, we show that the Elko-dominated universe can be completely determined by the composite scalar field $\phi=\dual{\lambda}\lambda$.
Since (\ref{Beom30}) is a third-order equation, it is easy to mistakenly think that the model suffers from Ostrogradsky ghost.
In this appendix, we will use a simple example to illustrate that even if the equation of motion of the composite scalar field is third-order, the model can be ghost-free, as long as there are no ghost modes in the original Lagrangian.

Consider a simple toy model whose Lagrangian is
\begin{equation}\label{appL1}
    L=\dot{\bar{z}}\dot{z}-V(\bar{z}z)~,
\end{equation}
where $z=x+iy$ is a complex variable, $\bar{z}$ is the complex conjugate of $z$, and the dot represents the time derivative.
After simple calculation, the Lagrangian $L$ can be transformed into
\begin{equation}\label{appL2}
    L=\dot{x}^{2}+\dot{y}^{2}-V(x^{2}+y^{2})~.
\end{equation}
Obviously, the Lagrangian (\ref{appL2}) is ghost-free, and the equations of motion can be obtained as
\begin{equation}\label{appeom1}
\ddot{x}+\frac{dV(x^{2}+y^{2})}{d(x^{2}+y^{2})}x=0~,~\ddot{y}+\frac{dV(x^{2}+y^{2})}{d(x^{2}+y^{2})}y=0~,
\end{equation}
If we are only concerned with the evolution of the composite scalar field $\phi=\bar{z}z=x^{2}+y^{2}$,
we can obtain from (\ref{appeom1}) that the equation of motion for $\phi$ is
\begin{equation}\label{appeom2}
    \dddot{\phi}+(4V_{\phi}+2V_{\phi\phi}\phi)\dot{\phi}=0~,
\end{equation}
where $V_{\phi}=\frac{dV}{d\phi}, V_{\phi\phi}= \frac{d^{2}V}{d\phi^{2}}$.
The equations of motion for the composite scalar field $\phi$ are third-order, but at the same time the Lagrangian (\ref{appL2}) clearly shows that the model is ghost-free.

This is exactly what happens in sec.~\ref{elkoFLRW}. Although the equation of motion for the composite scalar field $\phi=\dual{\lambda}\lambda$ is third-order, the equations of motion for the Elko field (\ref{Beom1})-(\ref{Beom2}) are second-order. Therefore, our model is free of Ostrogradsky ghost.

\bibliography{Bibliography}
\bibliographystyle{JHEP}

\end{document}